# The infinite red-shift surfaces of the Kerr and Kerr-Newman solutions of the Einstein field equations

*Gerald E. Marsh*

Argonne National Laboratory (Ret)
5433 East View Park
Chicago, IL 60615

E-mail: geraldemarsh63@yahoo.com

**Abstract.** In contrast to the Schwarzschild solution, the infinite red-shift surfaces and null surfaces of the Kerr solution to the axially-symmetric Einstein field equations are distinct. Some three-dimensional depictions of these surfaces are presented here for observers following the time-like Killing vector of the Kerr and Kerr-Newman solutions. Some similarities of the latter to the Reissner-Nordström solution are also discussed. In the case of the Kerr solution, the inner infinite red-shift surface terminates at the ring singularity. This is not the case for the Kerr-Newman solution where the infinite red-shift surface and the ring singularity have no points in common. The presence of charge severs the relation between the singularity and the infinite red-shift surface. This paper is also intended to fill a void in the literature where few, if any, adequate representations of the infinite red-shift surfaces and their relation to the singularity and horizons exist.

**Introduction**

The Kerr solution to the axially-symmetric Einstein field equations is generally accepted as corresponding to a rotating body. It is characterized by two parameters, the angular momentum per unit mass *a* and the mass *m*, and is generally discussed in Boyer-Lindquist coordinates [1]. These coordinates are very convenient for a variety of purposes, but make it difficult to visualize the actual shape of the infinite red-shift surfaces—to be defined below—as distinct from the event horizons corresponding to null surfaces that act as one-way membranes. For example, depictions of the relation of the inner infinite red-shift surface to the ring singularity are often glossed over or simply not shown as interest is generally centered on the ergosphere, the region between the horizon and the outer infinite red-shift surface, where the time-like Killing vector associated with the solution becomes space-like. Cartesian coordinates in Minkowski space, otherwise known as Kerr-Schild coordinates, offer a far more understandable representation of the infinite red-shift surfaces.

Observers at infinity follow the trajectories of the time-like Killing vector. Because the components of the metric tensor are independent of time, this Killing vector takes the form $\boldsymbol{\xi}_t$ = (1, 0, 0, 0). The $g_{00}$ component of the metric tensor is then given by $|\boldsymbol{\xi}_t|^2$. The Killing vector becomes null on surfaces where $g_{00} = 0$, but in the case of rotating black holes the surfaces themselves are not null and consequently not event horizons. The trajectories of $\boldsymbol{\xi}_t$ define the stationary frame—objects or observers moving along these trajectories appear to be stationary with respect to infinity. For these observers, the surfaces where $g_{00} = 0$ are infinite red-shift surfaces. Such observers are assumed in what follows.

For stationary metrics the infinite red-shift surfaces will be null surfaces only if $\boldsymbol{\xi}_t$ is hypersurface orthogonal—which, of course, it is not, since if it were the metric would be static. The infinite red-shift surfaces may still exist even when the event horizons do not.

In what follows, 3-dimensional representations of the infinite red-shift surfaces will be displayed for different, interesting values of the parameters *a* and *m*, along with the relation of these surfaces to the event horizons when they exist. The Mathematica® program used to generate these figures is given in the Appendix so that readers may generate their own figures for different values of the parameters. This is followed by a discussion of the infinite red-shift surfaces for the Reissner-Nordström and Kerr-Newman solutions.

**The Horizons and infinite red-shift surfaces**

Kerr-Schild coordinates were used by Kerr in one of the few faithful—albeit relatively forgotten—representations of the Kerr solution's infinite red-shift surfaces available in the literature [2]. At the time Kerr referred to these surfaces not only as the analogue of the Schwarschild sphere but also as null surfaces, but they are not null surfaces and therefore do not act as a one-way membrane.

Vishveshwara [3] used the following argument to show that the surface on which the Killing vector becomes null will be a null surface if and only if the rotation vector of the Killing vector field vanishes on the surface: define the normal vector, $n_a$, assumed not to vanish, on the family of hypersurfaces $\xi_a \xi^a$ = Const. defined by the Killing vector $\xi_a$ (indices range from 0-3). Then for stationary metrics

$$n^b n_b = \frac{1}{2}\left[\xi_a \xi^a \left(\xi_{b;c} \xi^{b;c}\right) - \omega_r \omega^r\right],$$

where

$$\omega^r = (-g)^{-\frac{1}{2}} \varepsilon^{rspq} \xi_s \xi_{p;q}.$$

For $n_a$ to be null when $\xi_a$ is null, $\omega_r$ must vanish. Now, the *r*-component of $\omega$ (called in various places in the literature the "rotation vector", the "angular velocity four-vector" and the "vorticity") is proportional to $\xi_{[s} \xi_{p;q]}$, whose vanishing implies the hypersurface orthogonality of $\xi_a$, in which case the metric would be static.

The infinite red-shift surfaces will be referred to here as null Killing surfaces to clearly distinguish them from the null surfaces corresponding to the horizons.

In Kerr-Schild coordinates, the Kerr solution is given by [4]

$$ds^2 = dx^2 + dy^2 + dz^2 - dt^2 + \frac{2m\rho^3}{\rho^4 + a^2 z^2}\left(k_\mu dx^\mu\right)^2, \tag{1}$$

where $k_\mu$ is the null vector field

$$k_\mu dx^\mu = dt + \frac{z}{\rho} dz + \frac{\rho}{\rho^2 + a^2}(xdx + ydy) + \frac{a}{\rho^2 + a^2}(xdy - ydx). \tag{2}$$

The surfaces of constant $\rho$ are confocal ellipsoids of revolution the equation for which is derived from the defining relations for oblate spheroidal coordinates:

$$x = a \cosh\xi \cos\eta \cos\phi$$
$$y = a \cosh\xi \cos\eta \sin\phi$$
$$z = a \sinh\xi \sin\eta.$$

Direct computation gives

$$\frac{x^2 + y^2}{a^2 \cosh^2\xi} + \frac{z^2}{a^2 \sinh^2\xi} = 1.$$

Setting $\rho^2 = a^2 \sinh^2\xi$ results in

$$\frac{x^2 + y^2}{\rho^2 + a^2} + \frac{z^2}{\rho^2} = 1. \tag{3}$$

$\rho$ is implicitly determined by this equation.

A ring singularity is located at $R := (x^2 + y^2)^{1/2} = a$ and $z = 0$ (where $\rho = 0$). This ring singularity bounds a surface having the character of a quadratic branch point in the complex plane; that is, if one passes through the surface from above (entering a region where, in Boyer-Lindquist coordinates, the coordinate labeling the oblate spheroidal surfaces of constant $r$ is negative) and were to loop around the ring singularity to again pass through the surface from above, one would return to the original starting space. The Kerr solution in the negative $r$ region is identical in structure to the positive $r$ part with $m$ being replaced by its negative.

To find the null Killing surfaces, one sets the $g_{00}$ component of the metric tensor in Eq. (1) equal to zero. The resulting equation for the null Killing surfaces is

$$\rho^4 - 2m\rho^3 + a^2 z^2 = 0 \tag{4}$$

Equation (3) is now solved for $\rho$ in terms of $R$ and $z$ and any one of the four solutions substituted into Eq. (4). The resulting equation is then solved for $z$ in terms of $m$, $a$, and $R$. There are eight solutions, only four of which are real. These correspond to four pieces of the null

*The infinite red-shift surfaces of the Kerr and Kerr-Newman solutions*

Killing surface. They can then be plotted together to obtain the 3-dimensional representation given in figures (1), (3), and (4).

The first example is for $m > a$ where the null Killing surfaces are distinct. This is shown in figure 1. Kerr's paper [2] shows a cross section of both this figure and figure 4. The horizons $r_\pm = m \pm (m^2 - a^2)^{1/2}$, where $r$ is the Boyer-Lindquist coordinate, are both located *between* the two displayed null Killing surfaces. The inner horizon $r_-$ is tangent to the inner null Killing surface at the two points equidistant from the origin on the $z$-axis where the surface intersects that axis; and $r_+$ is similarly tangent to the outer null Killing surface where it meets the $z$-axis (see figure 2(a)).

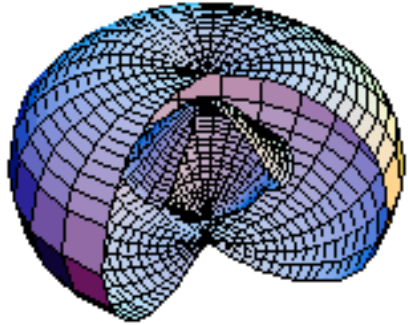

Figure 1. The Kerr null Killing surfaces for $m = 1.02\ a$. The ring singularity is at the cusp of the inner surface. The figure shows $z$, plotted as a function of $R$ for $\phi$ restricted to the range $0 \le \phi \le 3\pi/2$ so as to reveal the inner surface.

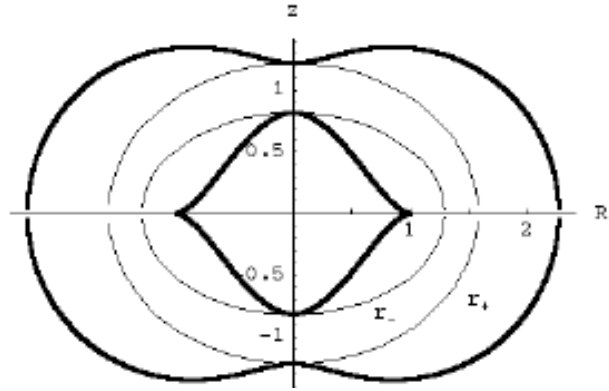

(a)

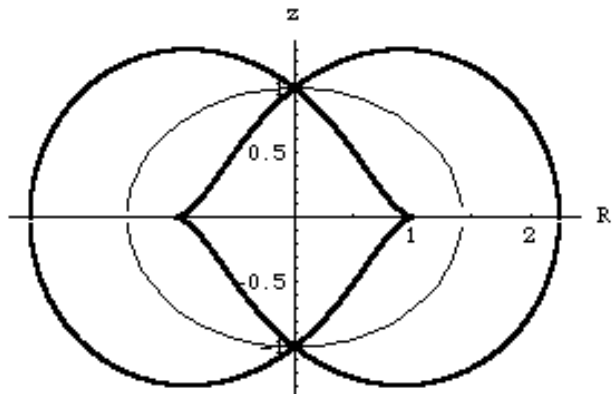

(b)

Figure 2. (a) The horizons $r_+$ and $r_-$ shown in relation to a cross-section of the null Killing surfaces of Figure 1; (b) The horizons $r_+$ and $r_-$ coalesce into a single horizon when $a = m$.

The case for $a = m$, with $m$ then set equal to unity is shown in figure 3. For this case, $r_+$ and $r_-$ coalesce into a single horizon located between the two null Killing surfaces. The horizon meets these surfaces at the two points equidistant from the origin on the $z$-axis where the null Killing surfaces also touch. The ring singularity, where the inner Killing surface terminates, remains hidden behind an horizon for this extreme case. This is shown in figure 2(b).

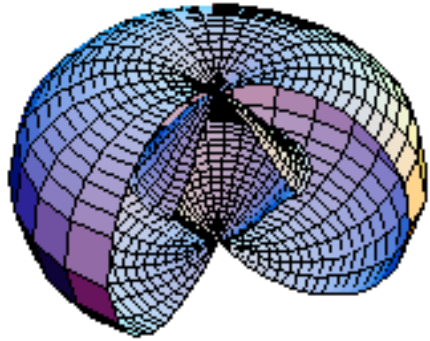

Figure 3. The null Killing surfaces of the Kerr solution for $a = m$. Note that these surfaces meet at two points equidistant from the origin on the positive and negative $z$-axis. The ring singularity remains at the cusp of the inner surface at $R = 1$ and $z = 0$.

The final case is that of $a > m$ where the null Killing surfaces open at the $z$-axis allowing the singularity at $z = 0$, and $R = 1$ to be seen from outside the surfaces. For this case, shown in figure 4, the horizons do not exist.

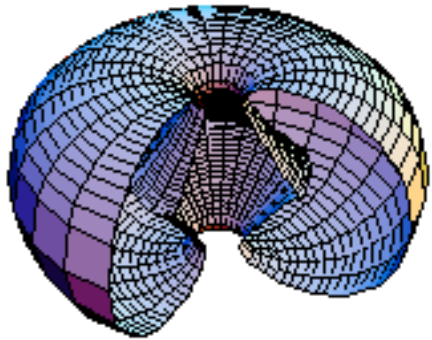

Figure 4. The Kerr null Killing surface for $a > m$. Here $m = 0.98\ a$. The ring singularity is again at the cusp of the inner part of the surface.

In all three cases, for plotting purposes, $a$ has been set equal to one.

*The infinite red-shift surfaces of the Kerr and Kerr-Newman solutions*For the case of $a > m$, where the horizons do not exist and the null Killing surfaces open up at the poles, the central region has the properties of a time machine [5]. This is a result of the fact that passing through the surface bounded by the ring singularity brings one into a region where the vector $\partial/\partial\varphi$ is time-like. In principle, one could cross the surface, travel in time by going around the *z*-axis, and then return through the surface and arrive at the starting point before the trip began.

Because of this bizarre behavior, many consider the case $a > m$, or others like it—to be discussed below, to be unphysical and believe that naked singularities will always be hidden behind event horizons. However, this *Cosmic Censorship* hypotheses has not yet been proven [6, 7, 8, 9, 10, 11].

The metric for the charged Kerr solution [12, 13], known as the Kerr–Newman solution, is similar to Eq. (1).

$$ds^2 = dx^2 + dy^2 + dz^2 - dt^2 + \frac{2m\rho^3 - e^2\rho^2}{\rho^4 + a^2 z^2} \left( k_\mu \, dx^\mu \right)^2, \tag{5}$$

where *e* is the charge, $\rho$ is again given by Eq. (3), and the null vector is the same as in Eq. (2). The ring singularity remains at $R = a$.

As before, the equation for the null Killing surfaces is obtained by setting $g_{00}$ in Eq. (5) equal to zero. This results in

$$\rho^4 - 2m\rho^3 + e^2\rho^2 + a^2 z^2 = 0. \tag{6}$$

This equation is somewhat more challenging to solve than Eq. (4). Again, Eq. (3) is solved for $\rho$ in terms of *R* and *z* and one of the four solutions substituted into Eq. (6). There are nine solutions to the resulting equation: one zero, four imaginary, and four that again correspond to pieces of the null Killing surfaces. Three dimensional renderings of these horizons do not add anything substantial to those already shown for the uncharged Kerr solution. However, for certain choices of the parameters *m*, *a*, and *e*, either one or both of the null Killing surfaces may not exist, and the relationship of the ring singularity to these surfaces changes.

Insight into the nature of the null Killing surfaces can be gotten by comparison with the Reissner-Nordström (charged Schwarzschild) solution, the metric for which is given by

$$ds^2 = -\left(1 - \frac{2m}{r} + \frac{e^2}{r^2}\right) dt^2 + \left(1 - \frac{2m}{r} + \frac{e^2}{r^2}\right)^{-1} dr^2 + r^2\left(d\theta^2 + \sin^2\theta \, d\phi^2\right), \tag{7}$$

where *e* is again the charge.

This metric generally has two null Killing surfaces, which are located at $r_\pm = m \pm (m^2 - e^2)^{1/2}$. Both are also null surfaces and consequently event horizons. For $e^2 = m^2$, there is only one horizon. Note that the gravitational effect of the charge on this spherically symmetric metric falls off as $1/r^2$ whereas that of the mass only as $1/r$. If $e^2 > m^2$, the metric has no horizons or null Killing surfaces but is non-singular everywhere except for the irremovable singularity at the origin [14]. The interesting thing about the singularity is that it is time-like so that clocks near the singularity run *faster* than those at infinity. For $e^2$ slightly greater than $m^2$, a clock approaching the radius where the horizon would be located for $e^2 = m^2$ slows down compared to those at infinity and then begins speeding up after it passes what would be the location of the $e^2 = m^2$ horizon until it reaches $r = e^2/2m$, where the metric takes the Minkowski flat-space form. A clock continuing to approach the singularity from this radius runs faster than one at infinity. This can be seen from figure 5.

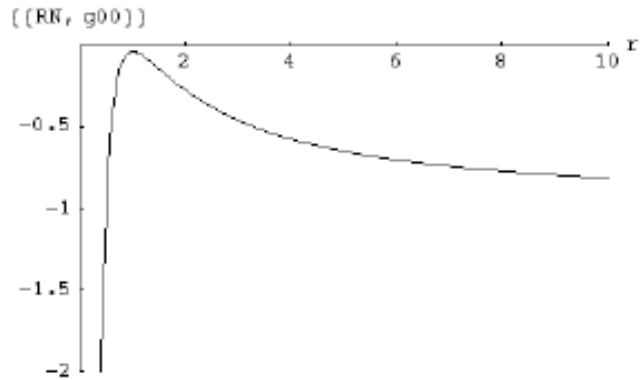

Figure 5. $g_{00}$ is shown as a function of $r$ for the Reissner-Nordström solution for $e^2 = 1$ and $m = 0.98$. Note that $g_{00}$ comes close but does not touch the $r$-axis indicating that no horizon or null Killing surface exists for this choice of the parameters $e$ and $m$.

The figure shows that, unlike the Schwarzschild solution, the singularity at $r = 0$ is time-like and repulsive—in that time-like *geodesics* will not reach the singularity.

In the case of the Kerr-Newman metric, consider the equatorial plane where $z = 0$. From Eq. (3), it is readily seen that for $z = 0$ the solution to this equation is $\rho = (R^2 - a^2)^{1/2}$. Equation (6), for $z = 0$, and the latter value of $\rho$ gives the following fourth order

*The infinite red-shift surfaces of the Kerr and Kerr-Newman solutions*

equation in *R* for the location of the intersection of the null Killing surfaces with the equatorial plane:

$$R^2 - a^2 - 2m\sqrt{R^2 - a^2} + e^2 = 0. \tag{8}$$

The relevant solutions to this equation are

$$R_+ = \left(a^2 - e^2 + 2m^2 + 2m\sqrt{m^2 - e^2}\right)^{\frac{1}{2}}, \tag{9}$$

$$R_- = \left(a^2 - e^2 + 2m^2 - 2m\sqrt{m^2 - e^2}\right)^{\frac{1}{2}}.$$

Similar to the case of the Reissner-Nordström solution, the Kerr-Newman metric has no null Killing surfaces for $e^2 > m^2$, but it also has none for $e^2 = m^2$ and $a > 0$, when the two intersection points with the equatorial plane coalesce. For this case, none of the solutions of Eq. (6) are real. An horizon exists for the Kerr-Newman metric only if $m^2 \geq e^2 + a^2$. For $a = 0$, the Kerr-Newman metric reduces to that of Reissner-Nordström.

For $e = 0$, the solutions of Eqs. (9) reduce to $R_- = a$ and $R_+ = (a^2 + 4m^2)^{1/2}$, corresponding to the Kerr solution, and for $a = 0$ we get the Schwarzschild result.

On the equatorial plane, $g_{00}$ for the Kerr-Newman solution is given by

$$g_{00} = \frac{2m\rho^3 - e^2\rho^2}{\rho^4} - 1. \tag{10}$$

Using $\rho = (R^2 - a^2)^{1/2}$, $g_{00}$ becomes

$$g_{00} = \frac{R^2 - a^2 - 2m\sqrt{R^2 - a^2} + e^2}{a^2 - R^2}. \tag{11}$$

If $a = 1$, and the other parameters are chosen to correspond to the Reissner-Nordström solution where $e^2 = 1$ and $m = 0.98$, neither of the intersection points of Eqs. (9) exist. The plot of $g_{00}$ looks essentially the same as figure (4), except that $g_{00}$ approaches negative infinity at $R = 1$, where the ring singularity is located, rather than 0 as was the case for the Reissner-Nordström solution. Had $m^2$ been chosen to be slightly greater than $e^2$, the intersection points of the null Killing surface would approach the point where the curve shown in figure 6 almost touches the *R*-axis near $R = 1.4$. Note that, like the Reissner-Nordström solution, the ring singularity at $R = 1$ is *time-like*.

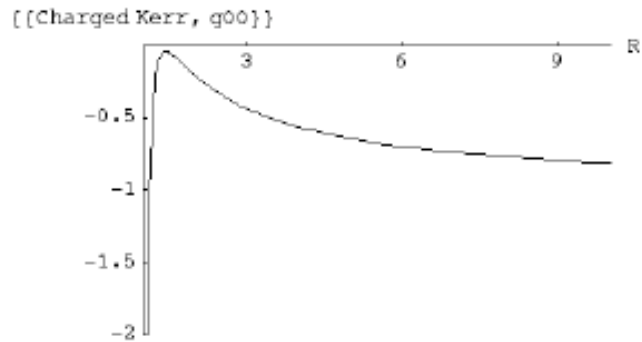

Figure 6. $g_{00}$ is shown on the equatorial plane as a function of $r$ for the Kerr-Newman solution for $a = 1$, $e^2 = 1$, and $m = 0.98$. Note that $g_{00}$ comes close but does not touch the *R*-axis indicating that no null Killing surface exists for this choice of the parameters $e$ and $m$. $g_{00}$ asymptotically approaches negative infinity as $R \to 1$, the location of the ring singularity.

One of the most interesting cases for the Kerr-Newman solution occurs when $a^2 + e^2 > m^2$ and $m > a > e$. The null Killing surface is then a toroid about the *z*-axis. The time-like Killing vector becomes space-like within this toroid. The relationship of the ring singularity to the null Killing surface is seen in figure (7) to significantly differ from that of the Kerr solution where at least one part of the surface terminates at the ring singularity. Here the ring singularity is *outside* the torroidal surface. The presence of charge not only makes the ring singularity time-like, but it severs the relation between the singularity and the null Killing surface.

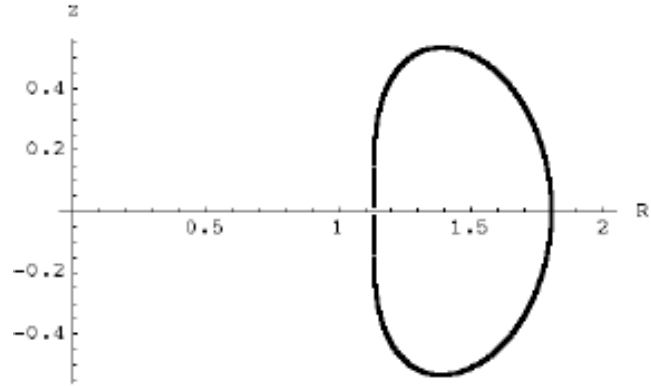

Figure 7. The null Killing surface for the Kerr-Newman solution with $m = 1.02$, $a = 1$, and $e = 0.9$ is obtained by rotating this figure about the $z$-axis. The time-like Killing vector becomes space-like within the toroid. The ring singularity is located at $R = 1$.

Even for the case $a^2 + e^2 = m^2$, where the ring singularity is enclosed by the null Killing surfaces, as well as an event horizon, the surface does not meet the ring singularity as it does in the case of the uncharged Kerr solution shown in Figure 1. This is shown in figure 8.

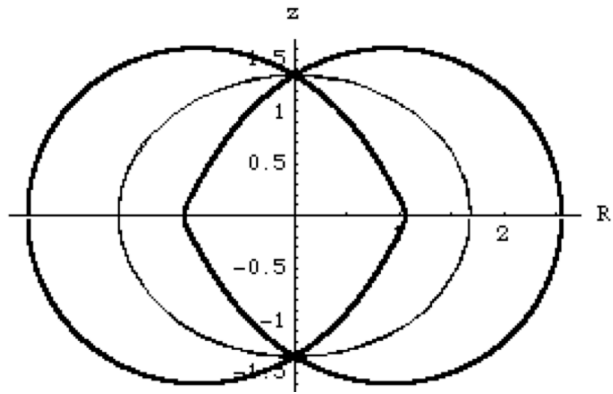

Figure 8. The null Killing surface for the Kerr-Newman solution with $m = 1.34536$, $a = 1$, and $e = 0.9$. Note that the inner surface toes not terminate at the ring singularity located at $R = 1$, but has an $R$ value, when $z = 0$, of about 1.05 (compare to Figures 1, 3, and 4). The single event horizon for $a^2 + e^2 = m^2$ is also shown (compare with figure 2(b)).

**Summary**

While an enormous literature has appeared since the discovery of the Kerr and Kerr-Newman solutions in the 1960s, the visualizations given here should add additional insight into these solutions. Perhaps the most interesting thing to come out of this study is the effect of the presence of charge on the null Killing surfaces when transitioning from the Kerr to Kerr-Newman solution. In the case of the Kerr solution the inner null Killing surface terminates at the ring singularity. When charge is added, this relationship is severed and the null Killing surfaces and ring singularity no longer have points in common. The singularity also becomes time-like. If the cosmic censorship hypothesis proves to be false, or have only limited applicability, the extreme Kerr or Kerr-Newman solutions could prove to have astrophysical implications.

The toroidal null Killing surface of figure 7, coaxial with the ring singularity but all parts of which lie at a greater radius than the singularity, is quite thought provoking. The torroid encloses the region of space-time where the time-like Killing vector becomes space-like.

The purpose of this paper is also pedagogical. Exact solutions to the Einstein field equations play an important role in the teaching of General Relativity. Of these, the solutions considered above—in addition to the Schwarzschild solution—are perhaps the most important. It is hoped that this material will found to be a useful complement the material found in most textbooks.

**Appendix**

In doing the calculations and plotting the null Killing surfaces for the Kerr solution I used Mathematica®, Version 3.0. While the program could be made more elegant, it will be relatively easy to modify in the form given below. The program for the 3-dimensional plots follows:

Solve[$\rho^4 + (a^2 - z^2 - R^2) \rho^2 - a^2 *z^2 == 0, \rho$];

$\rho$ /. %;

$\rho$ = Part[%, 1];

Solve[$\rho^4 - 2*m*\rho^3 + a^2 *z^2 == 0, z$]

(*The output cell from the above calculation should be converted in format to an input cell to avoid repeating the calculation for the a > m and m > a cases. Call the resulting input cell INPUT. The output from each reentry of this cell may be suppressed.*)

(*The case a = m = 1*)

% /. a -> m;

% /. m -> 1;

z /. %

h1 = %

Do[ f[i] = Part[h1,i], {i, 5, 8}];

r = {0, 0, 0, 0, 1, 1, 2.236067977, 2.236067977};

<<Graphics`ParametricPlot3D`

Do[ g[i] = CylinderPlot3D[ f[i], {R, 0, r[[i]]}, {phi, 0, 3Pi/2}, DisplayFunction -> Identity], {i, 5, 8}]

g[9] = Show[{g[5], g[6], g[7], g[8]}, DisplayFunction -> $DisplayFunction]

Show[%, Boxed -> False, Axes -> False]

(*The case a > m with a = 1*)

(*Enter INPUT*)

% /. m -> a/1.02;

% /. a -> 1;

z /. %

h2 = %

Do[ f[i] = Part[h2 ,i], {i, 5, 8}];

q = {0, 0, 0, 0, 0.2746197, 0.2746197, 0.2746197, 0.2746197}

r = {0, 0, 0, 0, 0.999, 0.999, 2.20105, 2.20105};

Do[ g[i] = CylinderPlot3D[ f[i], {R, q[[i]], r[[i]]}, {phi, 0, 3Pi/2}, DisplayFunction -> Identity], {i, 5, 8}]

g[9] = Show[{g[5], g[6], g[7], g[8]}, DisplayFunction -> $DisplayFunction]

Show[%, Boxed -> False, Axes -> False]

(*The case m > a with a = 1*)

(*Enter INPUT*)

% /. m  ->  1.02 a;

% /. a -> 1;

z /. %

h3 = %

Do[ f[i] = Part[h3 ,i], {i, 5, 8}];

r = {0, 0, 0, 0, 0.999, 0.999, 2.2719, 2.2719};

Do[ g[i] = CylinderPlot3D[ f[i], {R, 0, r[[i]]}, {phi, 0, 3Pi/2}, DisplayFunction -> Identity], {i, 5, 8}]

g[9] = Show[{g[5], g[6], g[7], g[8]}, DisplayFunction -> $DisplayFunction]

Show[%, Boxed -> False, Axes -> False]

(*Note: When plotting values for *a* and *m* other than those given above, if the plot ranges in the lists q and r are incorrectly determined the plots may have gaps or, if an imaginary number results, the plot command will not work. The correct numbers for the lists q and r that determine the plot ranges can be found by printing out f[i], where i = 5, 6, 7, 8 and then using the command % /. R -> (some appropriate number). The number can be varied so as to obtain a real values for the range of each of the f[i] that also gives a complete plot.  Approximate ranges can be determined using the command Do[T[i] = Table[{R, f[i]}, {R, 0, r[[i]], 0.01}], {i, 5, 8}] after entering trial lists with ranges great enough to cover the whole plot region. *)